\documentclass[]{acmart}
\AtBeginDocument{%
  \providecommand\BibTeX{{%
    \normalfont B\kern-0.5em{\scshape i\kern-0.25em b}\kern-0.8em\TeX}}}
\setcopyright{acmcopyright}
\copyrightyear{2024}
\acmYear{2024}
\acmDOI{XXXXXXX.XXXXXXX}

\usepackage{verbatim}
\usepackage{xcolor}

\begin{document}

\title{Automated Program Repair: Emerging trends pose and expose problems for benchmarks}

\author{Joseph Renzullo}
\authornote{Both authors contributed equally to this research.}
\email{renzullo@asu.edu}
\author{Pemma Reiter}
\authornotemark[1]
\email{pdreiter@asu.edu}
\affiliation{%
  \institution{Arizona State University}
  \streetaddress{727 E. Tyler St.}
  \city{Tempe}
  \state{Arizona}
  \postcode{85281}
  \country{USA}
}

\author{Westley Weimer}
\affiliation{%
  \institution{University of Michigan}
  \streetaddress{2260 Hayward St.}
  \city{Ann Arbor}
  \state{Michigan}
  \postcode{48109-2121}
  \country{USA}
}
\email{weimerw@umich.edu}

\author{Stephanie Forrest}
\authornote{Also at the Santa Fe Institute}
\affiliation{%
  \institution{Arizona State University}
  \streetaddress{727 E. Tyler St.}
  \city{Tempe}
  \state{Arizona}
  \postcode{85281}
  \country{USA}
}
\email{steph@asu.edu}

\begin{abstract}
Machine learning (ML) now pervades the field of Automated Program Repair (APR).
Algorithms deploy neural machine translation and large language models (LLMs) to generate software patches, among other tasks.
But, there are important differences between these applications of ML and earlier work. 
Evaluations and comparisons must take care to ensure that results are valid and likely to generalize.
A challenge is that the most popular APR evaluation benchmarks were not designed with ML techniques in mind.
This is especially true for LLMs, whose large and often poorly-disclosed training datasets may include problems on which they are evaluated.

\end{abstract}



\maketitle

\section{Introduction}

Automated Program Repair (APR), a subfield of software engineering, aims to reduce or eliminate direct human involvement in the repair of software defects, or bugs.
A variety of techniques have been developed, e.g.,  evolutionary computation~\cite{legoues2012GenProgGenericMethod, yuan2020ARJAAutomatedRepair}, methods incorporating templated mutation operators~\cite{liu2019TBarRevisitingTemplatebased}, semantic inference techniques~\cite{mechtaev2016AngelixScalableMultiline} targeting single-cause defects, and methods designed to handle multi-hunk bugs~\cite{saha2019HarnessingEvolutionMultiHunk}.
Increasingly, researchers have applied ML-based methods to APR tasks (Section~\ref{sec:s1-ml}), but data leakage is a concern (Section~\ref{sec:s2-data-leakage}).
Each new technique, or modification of an existing technique, tends to be developed by an independent research team, without reference to a common, formal definition of APR.
Benchmarks are not enough to standardize evaluation on their own (Section~\ref{sec:s3-benchmarks}).

As motivating examples, consider the following inconsistencies in the published literature:
\begin{itemize}
    \item \textit{Correctness}. VFix~\cite{xu2019VFixValueFlowGuidedPrecise} identifies correct patches that pass all test cases and are semantically or syntactically equivalent to the original bug-fix, while VRepair~\cite{chen2023NeuralTransferLearning} reports repair accuracy in terms of semantic equivalence to the original bug-fix, and SynFix~\cite{bhatia2018NeurosymbolicProgramCorrector} defines correctness simply as passing the test cases.
    Each of these is a reasonable definition, but collectively, their differences make it difficult to compare results.
    \item \textit{Fault Localization}. Sequencer~\cite{chen2019SEQUENCERSequencetosequenceLearning} assumes perfect fault localization at the line level, while iFixR~\cite{koyuncu2019ifixrbugreport} uses human-generated bug reports for fault localization; most earlier APR evaluations do their own fault localization.
    \item \textit{Termination criteria}. TBar~\cite{liu2019TBarRevisitingTemplatebased}, Angelix~\cite{mechtaev2016AngelixScalableMultiline}, and many other methods quit after finding one possible solution, while Dlfix~\cite{li2020DLFixContextbasedCode} stops when its fixed-length time budget is exhausted.
\end{itemize}

These are just a few examples of significant differences in how groups of authors report evaluation results on common benchmarks.
This paper provides a topical review of recent developments in the use of ML for APR tasks.
After a brief primer on commonly used terms (Section~\ref{sec:terms}), we consider the rise of diverse ML approaches as applied to APR (Section~\ref{sec:s1-ml}), specific concerns with respect to how training and testing data interact (Section~\ref{sec:s2-data-leakage}), and the role that benchmarks play in mediating and exposing these issues (Section~\ref{sec:s3-benchmarks}).
We discuss a new APR competition and a recent proposal to standardize ML model application to APR tasks (Section~\ref{sec:discussion}), before concluding the paper.

\subsection{Methodology}

Several high-quality general surveys of the APR field have been published~\cite{monperrusLivingReviewAutomated, gazzola2019AutomaticSoftwareRepair, legoues2019AutomatedProgramRepair}; here, we take a narrower view, examining all of the papers published at the top five software engineering venues since 2018 and focusing on ML in APR.
This allows us to analyze emerging trends in the venues favored by most APR researchers.
Methodologically, we analyzed Monperrus's The Living Review of Automated Program Repair (hal-01956501, version 5).
As it is frequently updated, Monperrus's living review includes a diversity of APR topics, including repair domains and error types, ancillary improvement methods, and human studies.
Using this review to seed our own, we determined the five venues publishing the largest volume of APR papers in The Living Review: ICSE (International Conference on Software Engineering), TSE (Transactions on Software Engineering), FSE (Foundations of Software Engineering), ASE (Automated Software Engineering), and EMSE (Empirical Software Engineering).
We then independently reviewed papers published in each of the five fora from January 1, 2018 through September 1, 2023.
We identified primary works for our study from paper metadata using the primary keywords \emph{patch} and \emph{repair}. 
We next searched for other potentially related works via secondary keywords: \emph{automate}, \emph{fix}, \emph{fault}, \emph{bug}, \emph{vulnerability}, and \emph{generate}.
Papers containing any of these terms (or their stems or variants) triggered a manual review of the abstract, which we examined for APR-related themes.
Of the 118 papers identified by our methodology, only 65 are cross-listed in the Monperrus living review.

\section{Background}
\label{sec:terms}


\subsection{Transformer Models and Attention}
Initially developed for NLP, neural machine translation (NMT) models employ neural networks for machine translation tasks like sentence prediction and translating from one language to another~\cite{cho2014PropertiesNeuralMachine}.
Current NMT models are usually implemented as encoder-decoder transformer architectures with attention~\cite{vaswani2017AttentionAllYou, luong2015EffectiveApproachesAttentionbased}, which improves on earlier LSTM-based~\cite{schuster1997BidirectionalRecurrentNeural} encoder-decoder architectures~\cite{cho2014PropertiesNeuralMachine}.
Transformer architectures vary: encoder~\cite{devlin2019BERTPretrainingDeep}, decoder~\cite{chen2021EvaluatingLargeLanguage}, and encoder-decoder pairs~\cite{raffel2020ExploringLimitsTransfer, lewis2020BARTDenoisingSequencetoSequence} are used in different circumstances.
Encoder components extract information and embed it in a latent space, while decoders generate content based on the information in the input's representation in that latent space.
Because encoder-only architectures have limited generative capabilities in zero-shot learning~\cite{wang2022WhatLanguageModel, tamborrino2020PretrainingAlmostAll}, decoder-only and encoder-decoder models are often adopted for generative tasks.
While not specific to transformers, \emph{attention} is a major architectural feature that assists scaling by accommodating input sequences of varying lengths~\cite{chaudhari2021AttentiveSurveyAttention}.
Attention allows the identification of input-output relationships when the position of those features is unknown.
This is important for tasks like APR, where inputs (usually tokenized source code) vary widely in their structure and length.

\subsection{Language Models}
Commonly used in NLP, language models are used to learn contextual representations of text and are usually implemented as deep neural networks.
A pre-trained language model (PLM) has parameters initialized from training on large corpora for generic text representation.
They amortize much of the effort required to train on such large datasets and can be fine-tuned later for specific tasks~\cite{tamborrino2020PretrainingAlmostAll, devlin2019BERTPretrainingDeep}.
Large language models (LLMs)~\cite{zhao2023SurveyLargeLanguage} are, as the name suggests, very large-scaled language models, usually with a billion or more parameters.
LLMs inherit the pre-training and fine-tuning paradigms and architectures of PLMs.
LLMs are seeing quick adoption in many fields, including APR, where their better performance at complex generative tasks is a relevant competitive advantage.


\subsection{Data Bias}
Data bias is a class of ML errors that occurs when features are not equitably represented in the dataset, leading to distorted results, compromised outcomes, and low accuracy~\cite{mehrabi2022SurveyBiasFairness}. 
Identifying data biases and their potential sources becomes increasingly difficult as the size of the training dataset increases. 
Unintended data bias is a common problem in natural language tasks when employing real-world data~\cite{borkan2019NuancedMetricsMeasuring}.
However, how bias may manifest in APR tasks is not well-studied. 
For example, an NPR model that uses Codex might be biased towards human-readable input-output pairs and be inaccurate when faced with algorithmically generated code, because Codex filters suspected auto-generated source code from its training corpora~\cite{chen2021EvaluatingLargeLanguage}.
While this and other potential code-related data biases should be explored in future research, researchers often mitigate known data bias by supplementing the training data with synthetic data~\cite{jaipuria2020DeflatingDatasetBias}.
Synthetic data are generated such that relevant statistical properties are conserved, while other factors are controlled.
The papers we reviewed did not explicitly discuss data bias or how to handle it.

\subsection{Data leakage}
In ML, data leakage occurs when the training data contains information or features relevant to the prediction.
This often leads to overestimating a model's likely performance on new, unseen data~\cite{kaufman2012LeakageDataMining}.
While data leakage can occur when the prediction or a proxy for the prediction is in the training data, other leakage and contamination occurs when individual cases contain similar content (near-duplication~\cite{barz2020WeTrainTest, laroca2023WeTrainTest, frobe2020SamplingBiasDue}) or the same content (duplication~\cite{lee2022DeduplicatingTrainingData}). 
ML researchers usually make a concerted effort when segmenting training from test data to avoid duplication, i.e., \emph{de-duplication} or \emph{decontamination}.
However, as we discuss in Section~\ref{sec:s2-data-leakage}, these methods are often insufficient.
In language models, duplication can occur at different levels of granularity or abstraction, i.e. sentences, documents, and datasets.
Similarly, duplication can occur at varying levels in source code, i.e., statements, functions, classes, and files.

\subsection{Beam Search}
In ML, particularly for NLP tasks, beam search~\cite{sutskeverSequenceSequenceLearning} determines how narrow or exhaustive the search is.
This is controlled by the hyperparameter \emph{beam size} $k$.
On each evaluative step, the $k$ sequences with the largest output weights from the set of remaining sequences are selected.

\section{Diverse ML approaches pervade APR}
\label{sec:s1-ml}

\begin{table}
    \centering
  \hspace{1em}
    \begin{tabular}{|c|c|}
\hline
keyword & count \\
\hline
program & 101 \\
repair+ & 93 \\ 
automated* & 36 \\
\checkmark learning & 25 \\
software & 25 \\
patch+ & 23 \\
automatic* & 19 \\
code & 15 \\
synthesis & 12 \\
\checkmark neural & 11 \\
\hline
  \end{tabular}
  \begin{tabular}{|c|c|}
\hline
keyword & count \\
\hline
generation & 10 \\
debugging* & 10 \\
\checkmark deep & 10 \\
analysis & 10 \\
testing & 9 \\
vulnerability* & 9 \\
\checkmark machine & 9 \\
fault* & 7 \\
empirical & 7 \\
execution & 7 \\
\hline
  \end{tabular}
  \begin{tabular}{|c|c|}
\hline
keyword & count \\
\hline
localization & 6 \\
based & 6 \\
test & 6 \\ 
java & 6 \\
engineering & 5 \\
bug* & 5 \\ 
symbolic & 5 \\
\dag overfitting & 5 \\
\dag correctness & 5 \\
fix* & 5 \\
\hline
  \end{tabular}
  \begin{tabular}{|c|c|}
\hline
keyword & count \\
\hline
quality & 5 \\
\checkmark networks & 4 \\ 
search & 4 \\
\dag assessment & 4 \\
static & 4 \\
mutation & 4 \\
fixing* & 4 \\
apr & 4 \\
bugs* & 4 \\
representation & 4 \\
\hline
  \end{tabular}
    \caption{Frequently used keywords for 118 APR papers: Author-selected keywords that were used four or more times in the dataset.  Keywords associated with ML are tagged with \checkmark and patch correctness with \dag.  Similarly, primary keywords are tagged with +, secondary with *.  }
    \label{tab:keywords}
\end{table}

Table~\ref{tab:keywords} summarizes the author-chosen keywords covered by the 118 papers in our corpus.
Of these, 29\% list ML-related keywords.
In 2018, this was relatively infrequent---fewer than one in six papers did so.
But by 2023, more than three in every four papers published employed ML techniques.
This is a striking change in a short period of time.

\subsection{The Rise of LLMs}

Increasing model size and data sizes of LLMs have led to a number of apparently emergent capabilities, including step-by-step reasoning and in-context learning~\cite{wei2023LargerLanguageModels, weiChainofThoughtPromptingElicits}.
These properties have intrigued researchers~\cite{bubeck2023SparksArtificialGeneral} and the public with the potential of LLMs as a component of artificial general intelligence (AGI).
But these properties, the factors that contribute to their emergence, and why they are absent in smaller models are not well understood~\cite{schaeffer2023AreEmergentAbilities}.
Regardless, LLMs are becoming popular with APR researchers, usually for patch generation.

\subsection{Code-Specific Language Models}

Some PLMs and LLMs include source code in their training corpora~\cite{chowdhery2023PaLMScalingLanguage, li2022CompetitionlevelCodeGeneration, scao2023BLOOM176BParameterOpenAccess, chen2021EvaluatingLargeLanguage}, often obtaining source code from public code repositories like GitHub or code-related help platforms like StackOverflow.
These models generate more accurate results in reasoning tasks when those tasks are formatted into code-like structures~\cite{madaan2022LanguageModelsCode}.
Specialized \emph{code} LLMs, fine-tuned on source code, have shown promise as generative models for tasks including code completion~\cite{chen2021EvaluatingLargeLanguage} and program synthesis~\cite{austin2021ProgramSynthesisLarge}.
Preliminary work applies LLMs to the problem of APR~\cite{jiang2021CURECodeAwareNeural, ahmed2023SynShineImprovedFixing, xia2022LessTrainingMore, fu2022VulRepairT5basedAutomated, ye2022SelfAPRSelfsupervisedProgram, xia2023AutomatedProgramRepair, fan2023AutomatedRepairPrograms, parasaram2023ReteLearningNamespace}.

\subsection{Architectures}
Other specialized ML architectures have been applied to code-related tasks.
When neural network techniques are used for APR, they are called \emph{neural program repair} (NPR).
While other code-related tasks have used structural architectures, like graph-based neural networks for code summarization~\cite{leclair2020ImprovedCodeSummarization}, most NPR architectures adopt NMT models. 
NMT-based NPR models can use custom implementations~\cite{tufano2019EmpiricalStudyLearning}, but code LLMs are increasingly popular in tools like AlphaRepair~\cite{xia2022LessTrainingMore} (CodeBERT~\cite{feng2020CodeBERTPreTrainedModel}), CoditT5~\cite{zhang2022CoditT5PretrainingSource} and RewardRepair~\cite{ye2022NeuralProgramRepair} (CodeT5~\cite{raffel2020ExploringLimitsTransfer}), and SelfAPR~\cite{ye2022SelfAPRSelfsupervisedProgram} (PLBART~\cite{ahmad2021UnifiedPretrainingProgram}).

In our review, we found that the transformer architecture is used more often than any other ML architecture. Of the 29 papers that use ML techniques, we identified 20 that use the transformer model, including the following 18 tools: RewardRepair~\cite{ye2022NeuralProgramRepair}, CURE~\cite{jiang2021CURECodeAwareNeural}, DLFix~\cite{li2020DLFixContextbasedCode}, 
VRepair~\cite{chen2023NeuralTransferLearning},
SeqTrans~\cite{chi2023SeqTransAutomaticVulnerability},
SynShine~\cite{ahmed2023SynShineImprovedFixing},
SequenceR~\cite{chen2019SEQUENCERSequencetosequenceLearning},
AlphaRepair~\cite{xia2022LessTrainingMore},
VulRepair~\cite{fu2022VulRepairT5basedAutomated},
Recoder~\cite{zhu2021SyntaxguidedEditDecoder},
Quatrain~\cite{tian2022ThisChangeAnswer},
SelfAPR~\cite{ye2022SelfAPRSelfsupervisedProgram},
TransRepair~\cite{li2022TransRepairContextawareProgram},
Reptory~\cite{namavar2022ControlledExperimentDifferent},
TENURE~\cite{meng2023TemplatebasedNeuralProgram},
Tare~\cite{zhu2023TareTypeAwareNeural},
Knod~\cite{jiang2023KNODDomainKnowledge},
and
Rete~\cite{parasaram2023ReteLearningNamespace}.
Other neural network-based architectures are used as classifiers, like the BiLSTM-based multi-classifier used in TRANSFER~\cite{meng2022ImprovingFaultLocalization}, as generative models, the RNN-based SynFix~\cite{bhatia2018NeurosymbolicProgramCorrector} and ASTNN-based code representation model used in AccPR~\cite{yang2021AcceleratingRedundancybasedProgram}, and finally as a learning model like the CNN used to learn features of fix patterns~\cite{liu2021MiningFixPatterns}.
The first wave of models to use ML for APR were largely NMT-based models, where the paradigm was to \textit{translate} from buggy code to repaired code.
The current trend is to adopt LLMs, where the paradigm is to delete a buggy line, and use generative text models to predict code to fill the resulting gap.


\subsection{Attention Mechanisms}
Attention is handled inconsistently in the transformer-based tools we considered: one does not mention a specific attention model~\cite{zhu2021SyntaxguidedEditDecoder}, five describe it in great detail~\cite{chi2023SeqTransAutomaticVulnerability, li2020DLFixContextbasedCode, chen2019SEQUENCERSequencetosequenceLearning, zhu2023TareTypeAwareNeural, jiang2023KNODDomainKnowledge}, while the remainder cite attention research like Vaswani~\cite{vaswani2017AttentionAllYou} and Luong~\cite{luong2015EffectiveApproachesAttentionbased}~\cite{jiang2021CURECodeAwareNeural, ding2020PatchingTranslationData, namavar2022ControlledExperimentDifferent, parasaram2023ReteLearningNamespace}, give a brief explanation~\cite{chen2023NeuralTransferLearning, li2022TransRepairContextawareProgram, meng2023TemplatebasedNeuralProgram}, or simply cite an inherited NMT model~\cite{ye2022NeuralProgramRepair, ahmed2023SynShineImprovedFixing, xia2022LessTrainingMore, fu2022VulRepairT5basedAutomated, tian2022ThisChangeAnswer, tian2020EvaluatingRepresentationLearning, ye2022SelfAPRSelfsupervisedProgram}.
Although CURE~\cite{jiang2021CURECodeAwareNeural} inherits its multi-headed attention architecture from its pre-trained language model GPT, it cites Vaswani's landmark paper~\cite{vaswani2017AttentionAllYou} for attention. 
In contrast, Seqtrans~\cite{chi2023SeqTransAutomaticVulnerability} describes attention in great detail while reusing OpenNMT which leverages Luong's attention mechanism~\cite{luong2015EffectiveApproachesAttentionbased}.
This variability of detail may result from the lack of accepted attention taxonomies~\cite{chaudhari2021AttentiveSurveyAttention}.

\section{Data Leakage contaminates models and benchmarks}
\label{sec:s2-data-leakage}


\subsection{Data Leakage}
Data leakage commonly occurs when training data includes information relevant to the task that it should not be able to access.
When the reviewed literature touched on data leakage, \textit{train-test segmentation} was common, i.e., the practice of distinctly separating the training corpus from the testing set such that coincident content is minimized.
SelfAPR~\cite{ye2022SelfAPRSelfsupervisedProgram}, which focuses on project-specific knowledge through self-supervised learning, splits its training and testing dataset along temporal boundaries, i.e., the timestamp of the bug-repairing commit in a project.
This approach is also used by CURE~\cite{jiang2021CURECodeAwareNeural}, which trains on data before 2006 in order to avoid contamination with the benchmarks (dating from 2007 onward).
Overfitting Detection System~\cite{ye2021AutomatedClassificationOverfitting}, a tool for correctness assessment of APR patches, adopts project-based benchmark segmentation to distinguish its training set (Bugs.jar~\cite{saha2018BugsJarLargescale}, Bears~\cite{madeiral2019BearsExtensibleJava}, and Defects4J~\cite{just2014Defects4JDatabaseExisting}) from testing by excluding from consideration any project (e.g. the Apache Math library) that exists in more than one of these sources, and therefore might foil standard random-sample segmentation due to duplicates.
For its syntactic training, RewardRepair~\cite{ye2022NeuralProgramRepair} leverages existing training corpora composed of bug-fix pairs and does not perform any train-test segmentation or decontamination for this content.
Synshine~\cite{ahmed2023SynShineImprovedFixing}, which uses a pre-trained RoBERTa transformer model, performed standard segmentation on its dataset consisting of 1.7M pairs of erroneous and fixed programs, selecting a random test set of 100K for evaluation.

\subsection{Contamination}
Contamination poses a significant risk to the validity of ML-based APR results, since it will tend to result in reported performance that fails to materialize in practice.
Duplicated data lead to bias in the trained models towards particular examples, which increases the likelihood that generated content is memorized~\cite{carlini2021ExtractingTrainingData}.
Our review did not find APR papers addressing the issue of near duplicates, though one did discuss potential contamination of PLMs~\cite{xia2023AutomatedProgramRepair}.
Decontamination can improve the performance of LLMs~\cite{chowdhery2023PaLMScalingLanguage, carlini2023QuantifyingMemorizationNeural}, but it is challenging.
First, since pre-trained weights are inherited, de-duplication of the training set by end-users is not possible.
Second, because details of training may not be available to the user, including the training corpus itself, decontamination of test data by outside users can be impossible.

We find that when LLMs are used in APR, five of the papers do not consider the impact of data leakage and contamination~\cite{ahmed2023SynShineImprovedFixing, xia2022LessTrainingMore, fu2022VulRepairT5basedAutomated, fan2023AutomatedRepairPrograms, parasaram2023ReteLearningNamespace}.
Xia, et al.~\cite{xia2023AutomatedProgramRepair} reports contamination information for each evaluated PLM with open-source training.
They enumerate the repairs that were identical to the developer provided change and were included in the training data. 
SelfAPR~\cite{ye2022SelfAPRSelfsupervisedProgram}, which opts for training its own LLM, removes all duplicate buggy samples from the training corpora.
CURE~\cite{jiang2021CURECodeAwareNeural}, which also retrains its LLM, populates its corpora with GitHub open-source Java projects rolled back before 2006, then manually removes duplicate bugs from their datasets, a coarse deduplication effort using the same methodology as CoCoNut~\cite{lutellier2020CoCoNuTCombiningContextAware}.
Comparing SelfAPR and CURE's decontamination policies, CURE's policy assumes independence among GitHub projects and between code commits and repairs (a strong assumption), where SelfAPR assesses each and retains only unique samples.
Three LLM methods for APR addressed data leakage by deduplicating their corpora.
One removed duplicates using syntax~\cite{le-cong2023InvalidatorAutomatedPatch}, another considered semantics~\cite{tian2022ThisChangeAnswer}, and the remaining did not specify its deduplication method~\cite{tian2020EvaluatingRepresentationLearning}.  

\subsection{Decontaminating Benchmarks}
Even when decontamination is an explicit goal, the capacity to accomplish it -- or even to quantify the level of contamination -- is often unavailable.
A standard decontamination method has not yet emerged for pre-trained models in language task settings, much less code task settings.
For example, GPT-3 used N-grams to identify contamination, where a \textit{gram} is a lowercase, whitespace delimited word without punctuation~\cite{brown2020LanguageModelsAre}. 
In comparison, GPT-4 uses multi-sample substring matching as a filter~\cite{openai2023GPT4TechnicalReport}.
Because its training corpora are unavailable, directly applying OpenAI's GPT decontamination policy to new benchmarks is infeasible for third parties using their models.
Additionally, standard approaches for estimating the likelihood of data contamination in training after the fact, such as quantifying the entropy gap between seen and unseen data~\cite{shejwalkar2021MembershipPrivacyMachine}, are not applicable when the model itself is not accessible (e.g., because users have API access).
\textit{Machine unlearning} updates a pre-trained model's existing weights, retraining or approximating the effects of removing implicated data~\cite{zhang2023ReviewMachineUnlearning, cao2015MakingSystemsForget, cao2018EfficientRepairPolluted, wu2022PUMAPerformanceUnchanged}.
Primarily focused on security and privacy concerns, research in machine unlearning is gaining momentum, particularly with the 2023 NeurIPS challenge announcement~\cite{NeurIPS2023Machine}.

\subsection{Memorization and Generalization}
\label{subsec:mem_gen}
Memorization, a type of data leakage, is defined as the propensity of language models to generate sequences from training data verbatim, without capturing their meaning~\cite{carlini2019SecretSharerEvaluating}.
Memorization may be beneficial for some language tasks that reproduce learned facts, but its benefit for code tasks is mixed.
It may help to repair bugs that are duplicates of what has been seen, but memorization is problematic when designing a model that is expected to generalize. 
While the memorization of specific training examples is difficult to prevent, recent research shows that removing duplicates or near-duplicates in the training corpora reduces verbatim memorization on average~\cite{lee2022DeduplicatingTrainingData}.

While a few language models outline a decontamination policy, even fewer provide a decontamination package with their model that detects contaminated test samples and produces a clean version of a benchmark, e.g., EleutherAI~\cite{phang2022eleutherai}.
Related work suggests two paths forward for tackling this issue.
First, testing approaches from language model security- and privacy-focused research can be leveraged to determine the likelihood of a test sample's existence in the dataset~\cite{carlini2021ExtractingTrainingData, peris2023privacy}.
One such method is the use of canary strings~\cite{carlini2019SecretSharerEvaluating}: easily detectable synthetic data that are obvious when reproduced.
Second, taking cues from adversarial text generation~\cite{gao2018black} and Seqtrans~\cite{chi2023SeqTransAutomaticVulnerability}, variable names and user-defined type names can be perturbed during validation, when a model is evaluated for generalization.

\subsection{Similarity and Granularity}
While document-, statement-, and sentence-level duplication is better studied in language models for non-code tasks, most de-duplication occurs at the file level for code models~\cite{allamanis2019nearduplicates, xu2022systematic}.
This may leave duplicate code excerpts intact, when duplication occurs at other levels of abstraction, e.g., for duplicated functions. 
As a simple example, consider multiple projects that reference a solution for a problem from StackOverflow, resulting in the adoption and adaptation of similar functions.
While lower-level segmentation may not be appropriate for code tasks like summarization, current NPR strategies that perform translation on small code fragments may be particularly susceptible to duplication. 
In our corpus, NPR techniques used various input granularities, ranging from max-token-limited context~\cite{ye2022NeuralProgramRepair, li2020DLFixContextbasedCode, chi2023SeqTransAutomaticVulnerability, ye2022SelfAPRSelfsupervisedProgram} to the entire buggy method~\cite{jiang2021CURECodeAwareNeural}, which determine the granularity of each model's semantic features.


\subsection{Impact of Near-Duplicates on Code Tasks}
Existing methods for deduplication often search for and remove exact matches.
This is insufficient for ML-based APR, especially in the context of emerging LLMs.
Near-duplicate code is a significant threat to language models~\cite{lopes2017dejavu}.
When near-duplicate code is included in training and benchmark datasets, performance metrics can inflate up to 100\% in comparison to de-duplicated corpora~\cite{allamanis2019nearduplicates}.
Language model contamination policies have leveraged metrics that measure similarities for natural language, rather than those used for code.
Luckily, the problem is amenable to text-based clone detection and code similarity metrics, which is often used to detect plagiarism~\cite{ragkhit2018codesimilarity} or license violations~\cite{golubev2020study}. 
While some string matching techniques can outperform specialized tools for code with heavy structural changes, highly specialized source code similarity detection techniques perform better than general textual measures of string similarity~\cite{ragkhit2018codesimilarity}.

\section{Benchmarks insufficiently standardize evaluations} 
\label{sec:s3-benchmarks}


\subsection{Performance Reporting}
Consider how results are reported for experimental evaluations using the ManyBugs benchmark~\cite{legouesManyBugsIntroClassBenchmarks2015}, an older collection of large, open-source C programs, compared to how they are reported for Defects4J~\cite{just2014Defects4JDatabaseExisting}, a 
more recent collection of small, open-source Java libraries.
For papers that report experiments on ManyBugs, authors usually report~\cite{mechtaev2016AngelixScalableMultiline} detailed run-time results in addition to the success rate of repairing defects.
However, papers evaluated on Defects4J frequently omit this information, focusing only on whether or not a candidate patch (or successful repair) was found.
When it is included, the reported information varies widely and can include an upper-bound on wall clock time~\cite{liu2019TBarRevisitingTemplatebased} or setting a limit on number of test suite evaluations~\cite{jiang2021CURECodeAwareNeural}, without reporting efficiency in sufficient detail to allow comparison across papers.



\subsection{Fault Localization}
An algorithm that has to handle fault localization is facing a different, harder problem, compared to an algorithm that does not.
Consider the distinction between \textit{patch generation} (the task of generating code modifications to be evaluated) and full \textit{automated repair} (an end-to-end process which includes the sub-problems of fault localization and a priority scheme when multiple solutions are proposed).
Some recent papers evaluate performance with and without assuming perfect fault localization~\cite{xia2022LessTrainingMore, zhu2021SyntaxguidedEditDecoder, fan2023AutomatedRepairPrograms, meng2023TemplatebasedNeuralProgram, zhu2023TareTypeAwareNeural} and  clearly specify which approach they have taken.
But performance reporting is not universal, which suggests that care must be taken when comparing results from different projects.

Some authors exclusively use information from running the test cases and tracing execution to inform a spectrum-based fault localization algorithm that predicts where the defect is likely to be~\cite{legoues2012GenProgGenericMethod, ye2022NeuralProgramRepair, gissurarsonPropRPropertybasedAutomatic2022, li2020DLFixContextbasedCode, saha2019HarnessingEvolutionMultiHunk, wenContextawarePatchGeneration2018, huaPracticalProgramRepair2018a, motwaniQualityAutomatedProgram2022, xuRestoreRetrospectiveFault2022, parasaramTridentControllingSide2021, chenContractBasedProgramRepair2021, afzalSOSRepairExpressiveSemantic2021, Oh2022pytertypeerrors, hua2018sketchfixlazycandidate, ye2022SelfAPRSelfsupervisedProgram, benton2020effectivenessunifieddebugging, ghanbari2019praprpracticalprogramrepairbytecodemutation, ginelli2022coderemovalpatches, aletiEAPRMappingEffectiveness2021, koyuncu2020fixminerapr, kim2019automaticpatchgenerationcontextbasedchange, li2023generatingconcisepatches}. Others take the location where the human applied the repair (provided by the benchmark for validating a solution) as the input directly, and focus attention on or around that location, excluding the problem of fault localization~\cite{jiang2021CURECodeAwareNeural, chen2023NeuralTransferLearning, chi2023SeqTransAutomaticVulnerability, chen2019SEQUENCERSequencetosequenceLearning, xia2022LessTrainingMore, fu2022VulRepairT5basedAutomated, yang2021AcceleratingRedundancybasedProgram, zhu2021SyntaxguidedEditDecoder, ding2020PatchingTranslationData, namavar2022ControlledExperimentDifferent, xia2023AutomatedProgramRepair, fan2023AutomatedRepairPrograms, meng2023TemplatebasedNeuralProgram, zhu2023TareTypeAwareNeural, jiang2023KNODDomainKnowledge, parasaram2023ReteLearningNamespace}.
NPR tools usually assume perfect fault localization, relying on the source file names and corresponding line numbers involved in the developer's repair to scrape the fault's location, rather than estimating it as part of the repair process~\cite{zhongStandUp4NPRStandardizingSetUp2022}. 
Between these extremes are a variety of choices, many of which have been adopted by APR algorithm designers~\cite{mechtaev2016AngelixScalableMultiline}.


\subsection{Correctness}
Many papers use a proxy for \textit{correctness}, most often comparing the generated patch to a single reference repair supplied by a human developer~\cite{meng2022ImprovingFaultLocalization, ye2022NeuralProgramRepair, jiang2021CURECodeAwareNeural, li2020DLFixContextbasedCode, liuEfficiencyTestSuite2020a, saha2019HarnessingEvolutionMultiHunk, xu2019VFixValueFlowGuidedPrecise, wenContextawarePatchGeneration2018, huaPracticalProgramRepair2018a, xuRestoreRetrospectiveFault2022, kechagiaEvaluatingAutomaticProgram2022, parasaramTridentControllingSide2021, chen2019SEQUENCERSequencetosequenceLearning, chenContractBasedProgramRepair2021, afzalSOSRepairExpressiveSemantic2021, ye2022SelfAPRSelfsupervisedProgram, zhongStandUp4NPRStandardizingSetUp2022, yang2022transplantfixgraphdifferencing, ghanbari2019praprpracticalprogramrepairbytecodemutation, ghanbari2019towardpracticalapr, namavar2022ControlledExperimentDifferent, ginelli2022coderemovalpatches, koyuncu2020fixminerapr, xia2023AutomatedProgramRepair, zhu2023TareTypeAwareNeural, jiang2023impactcodelanguagemodelsonAPR, motwani2023betteraprbugreports, jiang2023KNODDomainKnowledge, parasaram2023ReteLearningNamespace}.
This restricts the solutions that are judged acceptable to those that resemble the choices human developers happened to make, whether or not that is the only way (or even the best way) to satisfy the program's required functionality.
Since there are an infinite number of ways to implement any required functionality, defining correctness relative only to an arbitrary human example  is extraordinarily strict.


\subsection{APR Benchmarks Systematically Differ from ML Datasets}
Because of their different intended uses and construction requirements, APR benchmarks~\cite{just2014Defects4JDatabaseExisting, saha2018BugsJarLargescale, legouesManyBugsIntroClassBenchmarks2015, madeiral2019BearsExtensibleJava, linQuixbugsMultilingualProgram2017} are too small to train ML tools. 
They consist of a limited number of bug/fix pairs with developer-written patches and test cases to define constraints on expected program behavior.
The limited number of samples makes it difficult for ML models to be trained without other sources~\cite{ye2022NeuralProgramRepair}.
ML datasets, in order to be large enough, cannot be curated to the same extent that APR benchmarks are.
For example, the CrossVul dataset~\cite{nikitopoulos2021crossvul} is composed of differences between the commit for a bug fix and the previous program version.
However, these datasets do not include replicable build environments for patch evaluation, currently deemed unnecessary in ML patch generation~\cite{prenner2023runbugrun}.
Other large datasets of bug-fix pairs~\cite{jiang2021CURECodeAwareNeural} similarly lack execution environments like test suites, because they are essentially program-text-only.

\subsection{Access to source code, bug information, and repair information}
APR tools have requirements that interact with benchmark design.
For example, some APR tools require program-specific historical development information, using information retrieval (IR) methods that analyze a program's commit history or commit messages.
From our review, only iFixR~\cite{koyuncu2019ifixrbugreport} uses a bug-report based IR method.
These IR methods are supported in most benchmarks, since they provide links to each program's public code repository.
Full containment of this history within the benchmark is often infeasible.
Benchmarks that focus on small or novice-generated programs, like CodeFlaws~\cite{tanCodeflawsProgrammingCompetition2017} and IntroClass~\cite{legouesManyBugsIntroClassBenchmarks2015}, are exceptions, because the histories are short enough to be included.


\subsection{Reproducible program builds} 
Although earlier APR tools always relied on the ability to build a program from source code, until a recent NPR paper addressed this concern explicitly for ML evaluations~\cite{prenner2023runbugrun}, NPR techniques have avoided reproducing builds. 
Even with containerization, there can be conflicting requirements or dependencies that make building or running some APR tools incompatible with that program's container.
This may be particularly difficult for APR tools that augment build processes with their own requirements like Angelix~\cite{mechtaev2016AngelixScalableMultiline} and Prophet~\cite{longProphetAutomaticPatch2015}.
In comparison to C and C++ builds, Gradle and Maven are widely accepted build tools for Java programs that tightly control a program's external dependencies.
These standard build tools and the stability of Java and JRE make Java programs amenable for APR research.

From our review, we can see the impact of build overhead in the number of papers choosing to evaluate against ManyBugs~\cite{legouesManyBugsIntroClassBenchmarks2015}, an early C benchmark with incomplete build support, versus Defects4J~\cite{just2014Defects4JDatabaseExisting}, a Java benchmark complete with Maven build infrastructure.
There are five papers in our sample that contain data from evaluating ManyBugs~\cite{nollerTrustEnhancementIssues2022, parasaramTridentControllingSide2021, afzalSOSRepairExpressiveSemantic2021, cashinUnderstandingAutomaticallyGeneratedPatches2019, motwaniQualityAutomatedProgram2022}; however only two present new evaluation results, with the rest summarizing data of experiments from years past.
In contrast there are 44 papers in our sample that contain data from evaluating Defects4J~\cite{meng2022ImprovingFaultLocalization, ye2022NeuralProgramRepair, jiang2021CURECodeAwareNeural, li2020DLFixContextbasedCode, liuEfficiencyTestSuite2020a, saha2019HarnessingEvolutionMultiHunk, xu2019VFixValueFlowGuidedPrecise, wenContextawarePatchGeneration2018, huaPracticalProgramRepair2018a, chenProgramRepairRepeated2023, xuRestoreRetrospectiveFault2022, motwaniQualityAutomatedProgram2022, ye2022AutomatedClassificationOverfitting, chen2019SEQUENCERSequencetosequenceLearning, chenContractBasedProgramRepair2021, liu2021MiningFixPatterns, xia2022LessTrainingMore, wong2021varfixeditsearchexpressiveness, zhu2021SyntaxguidedEditDecoder, koyuncu2019ifixrbugreport, hua2018sketchfixlazycandidate, tian2022ThisChangeAnswer, ye2022SelfAPRSelfsupervisedProgram, yang2022transplantfixgraphdifferencing, wangAutomatedPatchCorrectness2020, tian2020EvaluatingRepresentationLearning, benton2020effectivenessunifieddebugging, cashinUnderstandingAutomaticallyGeneratedPatches2019, ghanbari2019praprpracticalprogramrepairbytecodemutation, ghanbari2019towardpracticalapr, yang2021wherewererepairingredientsdefects4j, aletiEAPRMappingEffectiveness2021, ye2021automatedpatchassessmentforprogramrepairatscale, koyuncu2020fixminerapr, kim2019automaticpatchgenerationcontextbasedchange, yu2019alleviatingpatchoverfitting, motwani2018doautomatedprogramrepairtechniquesrepairhardandimportantbugs, xia2023AutomatedProgramRepair, meng2023TemplatebasedNeuralProgram, zhu2023TareTypeAwareNeural, jiang2023impactcodelanguagemodelsonAPR, motwani2023betteraprbugreports, jiang2023KNODDomainKnowledge, le-cong2023InvalidatorAutomatedPatch}.
While there are other factors that influence the differences in adoption patterns for these two datasets, e.g. the fact that they target different languages, some authors~\cite{lutellier2020CoCoNuTCombiningContextAware} have been unable to execute ManyBugs and so performed syntactic comparison to the human-supplied repair as a proxy for evaluation.

While obviously at a disadvantage due to this build complexity, C and C++ programs account for a significant fraction of software in development~\cite{bissyandePopularityInteroperabilityImpact2013}.
This prevalence, together with the emergence of code LLMs for multiple languages~\cite{cassano2023polyglotneuralbenchmark}, indicates that APR benchmarks would benefit from a consolidated build and run environment that enables a variety of APR tools and language bases.

\subsection{Consistent reporting measures}
We found that NPR techniques have not converged on a standard metric for search efficiency and generate a wide number of unverified patches for each buggy program.
For example, unlike traditional APR tools which guarantee plausibility for each patch at a minimum, NPR techniques generate patch candidates (beam search), which are checked afterward for plausibility and correctness.
We found hints that beam size is trending as a standard, with justification of beam size by reference to other evaluations~\cite{jiang2023KNODDomainKnowledge, chen2023NeuralTransferLearning}.  
Beam sizes ranged anywhere from 50~\cite{chi2023SeqTransAutomaticVulnerability, chen2023NeuralTransferLearning} to 200~\cite{ye2022NeuralProgramRepair} to 1000~\cite{jiang2023KNODDomainKnowledge, parasaram2023ReteLearningNamespace}. 
However, beam size is an ineffective measure of search efficiency as it neglects to account for the model's generative vocabulary size and sequence length.
Unfortunately, we found that generative vocabulary size was not always reported~\cite{jiang2023KNODDomainKnowledge, li2020DLFixContextbasedCode}, making search efficiency comparisons difficult.
When vocabulary is accounted for, NPR APR models have widely varying effective search sizes:
SeqTrans ($k=[1...50]$, $|Y|=8,000$);
VRepair ($k=50$, $|Y|=5,000$);
RewardRepair($k=[30,100,200,500]$, $|Y|=32,128)$);
CURE($k=[1000]$, $|Y|=50,000$).

\section{Discussion}
\label{sec:discussion}

\subsection{Standup4NPR}

StandUp4NPR~\cite{zhongStandUp4NPRStandardizingSetUp2022} was introduced to standardize the evaluation of ML-based APR tools.
It has a well-defined interface for training and testing, and can impose resource restrictions on all models to facilitate accurate comparison between tools.
The training and testing data are taken from existing datasets and benchmarks, and are then subsequently processed and filtered, in an attempt to minimize the incidence of exact duplicates.

Many aspects of this project are promising: standardizing data access, resource usage, runtime, and evaluation criteria across a variety of models creates a much more level field for comparison.
However, other design features of StandUp4NPR can pose problems.
By restricting data access to the training examples provided, what is being tested is how efficient and effective an architecture is at leveraging the data in the training dataset.
However, it rules out other tasks that can improve APR efficiency of ML tools: careful curation of other, independent datasets; using auxiliary inputs (like program history, or the history from other projects) not available within the benchmark, etc.

A related issue is that, because StandUp4NPR was formulated for the evaluation of neural techniques, it is at times ill-designed for use by traditional APR algorithms.
GenProg~\cite{legoues2012GenProgGenericMethod} and ARJA~\cite{yuan2020ARJAAutomatedRepair}, for example, use information about how a mutation has performed to affect what they do at the next time-step.
But the StandUp4NPR framework assumes that all patches are generated simultaneously and then evaluated independently.
This lack of an evaluation feedback loop means that many search-based techniques cannot be compared under this framework.

\subsection{APR-COMP 2024}

Recently, the 1st International Competition on Automated Program Repair was announced.
It has four competition tracks: functional errors, vulnerabilities, student assignments, and AI-generated code.
For the functional repair track, benchmark programs are sampled from existing benchmarks~\cite{legouesManyBugsIntroClassBenchmarks2015, just2014Defects4JDatabaseExisting, madeiral2019BearsExtensibleJava, saha2018BugsJarLargescale}.
This means that it inherits the issues that these benchmarks have, and that the content is knowable to model designers in advance.
This differs from how some annual ML competitions, e.g. ImageNet~\cite{dengImageNetLargeScaleHierarchical}, are structured; they released a public training dataset, but held in private the test data.
That said, having a standard API for access to multiple benchmarks is a large benefit for the interoperability of repair tools.
Also, as with StandUp4NPR, common limits on evaluation time and resource usage facilitate comparison of the resulting data.
Finally, because submission is via public pull request, it is in practice requiring disclosure of an executable object, which not all APR papers contain.

\section{Conclusion}
APR is a vibrant and fast-evolving field within software engineering.
There is already economic benefit from the application and deployment of APR techniques, evidenced by its increasing uptake by industry.
However, as it has matured, the field has concentrated evaluation effort around a few popular benchmarks, which are now showing their age.
In particular, the rise of ML approaches in APR poses problems for the evaluation of existing benchmarks.

In this paper, we have detailed the origin of some of these issues, as well as provided ways we might translate insights from other fields to tackle the challenges we have in common.
Data leakage is a significant threat to the validity of results, particularly language models whose training corpora are unavailable or undisclosed.
The wide variety in the reporting of results additionally makes comparing across evaluations challenging at best and infeasible at worst.
Due to the issues that are inherent to the methods we use to evaluate the success of APR tools, more and larger empirical studies and comparisons will not shed light until these problems are addressed.

\bibliographystyle{ACM-Reference-Format}
\bibliography{review}

\end{document}